# Probably Something:
# A Multi-Layer Taxonomy of
# Non-Fungible Tokens

Eduard Hartwich[1], Philipp Ollig[2], Gilbert Fridgen[1], Alexander Rieger[1]

[1] University of Luxembourg, Interdisciplinary Centre for Security, Reliability, and Trust (SnT)
[2] University of Bayreuth, FIM Research Centre

## Abstract

**Purpose** – This paper aims to establish a fundamental and comprehensive understanding of Non-Fungible Tokens (NFTs) by identifying and structuring common characteristics within a taxonomy. NFTs are hyped and increasingly marketed as essential building blocks of the Metaverse. However, the dynamic evolution of the NFT space has posed challenges for those seeking to develop a deep and comprehensive understanding of NFTs, their features, and capabilities.

**Design/methodology/approach** – Utilizing common guidelines for the creation of taxonomies, we developed (over three iterations), a multi-layer taxonomy based on workshops and interviews with 11 academic and 15 industry experts. Through an evaluation of 25 NFTs, we demonstrate the usefulness of our taxonomy.

**Findings** – The taxonomy has four layers, 14 dimensions and 42 characteristics, which describe NFTs in terms of reference object, token properties, token distribution, and realizable value.

**Originality** – Our framework is the first to systematically cover the emerging NFT phenomenon. It is concise yet extendible and presents many avenues for future research in a plethora of disciplines. The characteristics identified in our taxonomy are useful for NFT and Metaverse related research in Finance, Marketing, Law, and Information Systems. Additionally, the taxonomy can serve as an information source for policymakers as they consider NFT regulation.

**Paper Type** Research Paper

# Introduction

Considerable hype has developed around the Metaverse vision (Kim, 2021). The metaverse refers to a three-dimensional digital world in which users can interact and collaborate in real-time, thus offering a new way to be online (Kim, 2021; Stackpole, 2022). Attention to this vision has increased after Facebook's announcement to change its widely known company name to 'Meta' and refocus its mission to "bring the metaverse to life" ("The Facebook Company Is Now Meta," 2021). However, the idea of virtual worlds, including a virtual landscape and "citizens", is not new. Early examples, such as Second Life, have been around since the early 2000s (Berente *et al.*, 2011; Chandra and Leenders, 2012). Unlike purely virtual worlds and traditional online environments (Haines, 2021; Sjöblom *et al.*, 2020), however, the Metaverse is expected to feature a network of shared virtual environments in which users can immerse themselves, interact, and blur the line between virtual and real life (Kim, 2021).

With the development of metaverse platforms and applications picking up pace, researchers are beginning to identify various directions for relevant research. One such direction are non-fungible tokens (NFTs) and their role in the metaverse (Dwivedi *et al.*, 2022). NFTs allow to identify holders of certain rights associated with unique digital or physical reference objects (Dowling, 2022b; Regner *et al.*, 2019). Although they do not embody rights to these reference objects in a narrow legal sense (Chalmers *et al.*, 2022; Fairfield, 2021), NFTs enable efficient ways to digitally signal who holds these rights at any given time (Rendle and McLean, 2021; Wang *et al.*, 2021). This can make them an important tool for interlocking the physical and virtual worlds (Dwivedi *et al.*, 2022).

NFTs allow to reference various objects, such as digital artworks, collectibles, permits, certificates, real estate, and a variety of accessories (Dowling, 2022; Pawelzik and Thies, 2022). Because of the dynamic evolution of the NFT space, it is difficult to keep track of the change and innovations, and most users and organizations still struggle to identify, use and implement NFTs (Wang *et al.*, 2021). For example, two-thirds of people in Germany have never heard of NFTs (Streim and Faupel, 2022); in the US, this number is even lower, as only 1 in 4 adults know what an NFT is (Glum, 2022). Simultaneously, many consumer brands such as adidas, Gucci, and Coca-Cola have launched NFT collections or are selling NFTs on metaverse platforms (Kim, 2021). Furthermore, crypto trading platforms such as Binance, Coinbase, and Kraken have introduced their own NFT marketplaces, opening the trading of NFTs to millions of users. Following these developments, we identify the need for greater common understanding, the use of common language, and an overview of the possible characteristics of NFTs which could be used by those that wish to engage with NFTs. Hence, we address the following research question:

*How can Non-Fungible Tokens be classified?*

To address this question, we set out to identify similarities, comparability, and differences between NFTs. We follow an iterative development method (Kundisch *et al.*, 2021; Nickerson *et al.*, 2013) and create a multi-layer taxonomy for NFTs. Our taxonomy consists of 42 characteristics clustered in 14 dimensions, aggregated in 4 layers. We developed it over three iterations relying on a selection of NFT projects, as well as using exploratory workshops and interviews with 11 academic and 15 industry experts. We evaluated the taxonomy by classifying 25 exemplary NFTs, out of which we showcase two in this paper, to demonstrate its usefulness. As a result, this paper provides both theoretical and practical contributions that lay a solid foundation to explore and structure future use cases (e.g., in the Metaverse).

This paper is laid out as follows. In Section 2, we provide a theoretical background on key concepts used throughout the paper, namely Metaverse, Cryptographic Tokens, Non-Fungible Tokens, and NFTs in management and law. Next, we introduce the research method (Section 3) that we applied to develop the Multi-Layer Taxonomy presented in Section 4. In Section 5, we evaluate the taxonomy using 'real-world' NFT projects before discussing our work's theoretical, and practical implications (Section 6). Lastly, we conclude by summarizing our results, identifying limitations, and providing an outlook on potential future research (Section 7).

# Theoretical Background

## Metaverse

The Metaverse refers to a digital world in which users can interact and collaborate in real-time (Kim, 2021). Although virtual worlds are not new (as the example of Second Life shows (Berente *et al.*, 2011; Chandra and Leenders, 2012)), digital technologies such as virtual or augmented reality continue to blur the line between the physical and the digital. This distinguishes it from traditional online environments (Haines, 2021; Sjöblom *et al.*, 2020). Due to the emergence of new technologies, the Metaverse is supposed to be more immersive not only for gaming, but also for shopping or attending tele-health-appointments, live education courses, and other events.

Following these developments, a host of early adopters have begun to explore opportunities of the Metaverse vision. To emphasize its belief, Facebook renamed itself Meta (Isaac, 2021), which made it the biggest and potentially most prominent player in the Metaverse discourse. While Meta is focusing on creating its own Metaverse platform, other companies such as Gucci, adidas, and Chipotle have started to explore options to offer digital products and digital experiences within a potential Metaverse economy. In line with this evolution, researchers are continuously identifying interdisciplinary research directions including aspects of psychology, culture, security and economics (Dwivedi *et al.*, 2022).

Against this backdrop, cryptographic and non-fungible tokens are often discussed as an integral part of 'the Metaverse' (Dwivedi *et al.*, 2022). These tokens allow individuals to own the likes of virtual land, buildings, equipment, avatars, clothing, or vehicles (NonFungible Corporation, 2022). In addition, ownership of such tokens grants access to private communities, voting rights, and potential business models such as play-to-earn games (Stackpole, 2022).

## Cryptographic Tokens

In October 2008, Nakamoto (2008) introduced Bitcoin as the world's first cryptocurrency. A cryptocurrency is a cryptographic token that, much like regular currency, can be used to pay for goods or services. It relies on cryptographic primitives and distributed ledger technologies (DLTs) to verify transactions and ensure their authenticity (Farell, 2015). To make a transaction, users typically employ digital wallets (also called crypto-wallets) that store public-private keypairs. The public key serves as their address in the cryptocurrency network and the private key permits identification and authentication of transactions (Alt, 2022; Sedlmeir *et al.*, 2021). Once a new transaction is submitted to the network, it is verified and added to a continuously growing distributed ledger of transactions that ensures tamper-resistant recording (Butijn *et al.*, 2020).

Based on Nakamoto's ideas, other developers started designing alternative DLTs and experimenting with use cases for cryptographic tokens beyond pure monetary transactions. Ethereum – one of the earlier successors of Bitcoin and as of writing the second largest cryptocurrency by market capitalization (coinmarketcap.com, 2022) – pioneered the automated execution of more complex programming logic by the blockchain network using so-called smart contracts (Butijn *et al.*, 2020; Wood, 2014). Smart contracts are algorithms that reside on a distributed ledger and allow for the design of more sophisticated cryptographic tokens. (Christidis and Devetsikiotis, 2016).

While the Bitcoin token represents a so-called payment token, Ethereum enabled the development of new token types (namely security and utility tokens) and token standards (Di Angelo and Salzer, 2020). A token standard describes a set of rules that defines how a token must be designed to interact and interface with the underlying DLT protocol and ensure the composability of smart contracts (ethereum.org, 2022). The first token standard that Ethereum introduced in 2015 was the *Ethereum Request for Comment* 20 (ERC-20), which set the standard for fungible tokens designed on the Ethereum blockchain (crypto.com, 2022), followed by ERC-721 and later ERC-1155 both of which put non-fungible token standards on the map (Di Angelo and Salzer, 2020; Martin and Kellar, 2021). Ever since, the NFT hype led other distributed ledger networks to develop their own token standards and enable NFTs native to their network, e.g., Solana (solanart.io, 2022) or the Binance Smart Chain (binance.com, 2022).

## Non-Fungible-Tokens

Fungibility describes the interchangeability and replicability of an object with an "identical or similar object" (U. W. Chohan, 2021). For instance, a $10 bill can be exchanged for another $10 bill or even two $5 bills. Other examples include commodities or common shares in a company – one stock certificate is just as good and valuable as another. In contrast, "*non-fungible*" describes unique objects that cannot be substituted, divided, or replaced (Nadini *et al.*, 2021). Much like fungibility, the concept of non-fungibility is neither new nor exclusive to the DLT or Metaverse space. Real-world examples of non-fungible objects are unique paintings such as the Mona Lisa, real estate, or diamonds (Pawelzik and Thies, 2022). Each of these objects has unique properties and cannot be replaced. An apartment in New York cannot be substituted for an apartment in Berlin or London.

While it is fairly straightforward to track rights associated with non-fungible objects in the physical world, this is much harder in the digital world because digital objects can be replicated (Kugler, 2021). This challenge is one to which NFTs are seen as a potential solution (Chalmers *et al.*, 2022). To create an NFT, a user must create a token and record it on a distributed ledger through a process called minting. During the minting process, the creator of the NFT sends data pertaining to the NFT to a smart contract that creates a token following a certain standard. The smart contract then processes the request and mints (creates) the NFT, thus tying the NFT to an immutable address in the network (Wang *et al.*, 2021). By following a 'unique token' standard such as ERC-721, the NFT creator can mint a single, unique token, meaning there will only ever be one issue of this NFT. Arguably the most prominent example for an ERC-721 token is Beeple's artwork "Everydays: The first 5000 days" (Christie's.com, 2021). In contrast, ERC-1155 tokens enable multiple issues of a token. That is, an NFT creator can decide how many issues of his NFT can exist (Kuhn *et al.*, 2021). For example, the "adidas Originals: Into the Metaverse (Phase 1)" NFT is created as ERC-1155 token counting 30,000 issues.

While NFTs have been around for almost a decade, they gained substantial traction in 2021 in response to Beeple's record NFT sale and the general increase in cryptocurrency market capitalization. At the end of 2021, global NFT market capitalization temporarily peaked at USD 41 billion (Dailey, 2022), and "NFT" became the most used search term on Google (İkiz *et al.*, 2022). This hype is currently centered around art and gaming-related NFTs. However, NFTs have a much broader spectrum of use cases ranging from art to real estate to virtual land, or to a specific set of rights (Arslanian and Fitzgerald, 2021).

## *NFTs in management and law*

In line with this broad set of use cases, various disciplines have started to study NFTs. For example, Finance and Marketing examine NFTs from pricing (Borri *et al.*, 2022; Dowling, 2022a; Xia *et al.*, 2022) or branding perspectives (R. Chohan and Paschen, 2021). Legal research, on the other hand, studies matters of ownership, copyrights, or trademarks (Murray, 2022), as well as patents and intellectual property (Bamakan *et al.*, 2022; Okonkwo, 2021).

Scholars in Finance have responded to the increase in market capitalization, in addition to partially dizzying valuations, by raising questions about the pricing of NFTs (Dowling, 2022a), their potential as an investment vehicle, and the role of NFTs in relation to other financial asset classes (Borri *et al.*, 2022; Mekacher *et al.*, 2022; Xia *et al.*, 2022). In line with the NFT market's current focus on art (İkiz *et al.*, 2022), others have begun to examine price determinants in NFT art markets (Horky *et al.*, 2022) and the potential roles of NFTs in creative industries (Chalmers *et al.*, 2022).

Marketing scholars, in turn, have reacted to many brands embracing NFTs for marketing purposes, and they explore how NFTs can be used as marketing tools (Hofstetter *et al.*, 2022). This, in turn, led others to analyze the marketing implications of NFTs more broadly and ways of involving them in customer interaction (R. Chohan and Paschen, 2021) . Colicev (2022), for instance, explores how NFTs can be used in community building and how companies can best leverage NFTs to strengthen their brands.

Lastly, legal scholars have started to fit NFTs into current legal frameworks. For instance, Murray (2022) argues that the NFT and the underlying reference object can be two distinctly different legal objects that grant different rights to their respective owners, for example, copyrights or trademarks. Other examples include the implications of NFTs for intellectual property protection and commercialization (Bamakan *et al.*, 2022; Okonkwo, 2021).

Research in all these fields provides important insights into the NFT phenomenon. However, these dispersed discussions make it difficult to gain a holistic picture of NFTs, their characteristics, and their capabilities. To sharpen the understanding of NFTs and to aid future research in defining their scope more precisely, we next analyze how NFTs can be classified by following a structured, methodical approach.

# Research Method

To create a structure to better understand the opaque field of NFTs and to establish a consistent nomenclature, we developed a multi-layer taxonomy, following recommendations by Nickerson *et al.* (2013) and Kundisch *et al.* (2021). Taxonomies aim to structure and systematize complex emerging domains where little knowledge is available (Nickerson *et al.*, 2013). They are a useful means to provide structure to complex and messy data as they assist in the understanding and analysis of different concepts and their characteristics (Berger *et al.*, 2020), thus laying a foundation for future research. Our taxonomy is intended for use by IS scholars, NFT creators, NFT buyers, and policymakers that consider NFT regulation.

Taxonomies have been created and used successfully as foundational elements in various related areas, such as for the "understanding of blockchain-based tokens" (Oliveira *et al.*, 2018), "blockchain-enabled crowdfunding (ICOs)" (Fridgen *et al.*, 2018) or "the impact of blockchain technology on business models" (Weking *et al.*, 2020). Leaning on this research, we adapted the iterative taxonomy development process proposed by Nickerson *et al.* (2013). This process follows seven steps: (1) Determine meta-characteristics, (2) determine ending conditions, (3) decide on an empirical-to-conceptual and/or conceptual-to-empirical design approach, (4) identify or conceptualize objects, dimensions, and characteristics, (5) examine objects, (6) create the taxonomy, and lastly (7) check if previously determined ending conditions are met. If not, repeat steps 3-7 until they are met.

Following this process, we first defined "design parameters and characteristics of Non-Fungible Tokens" as our meta-characteristic. Second, we defined objective and subjective ending conditions. Objective ending conditions ensure that a taxonomy fulfills objectively all necessary and essential criteria (Kundisch *et al.*, 2021). They assure that a taxonomy has taken account of a representative sample of the underlying object and that it "consists of dimensions each with mutually exclusive and collectively exhaustive characteristics" (Nickerson *et al.*, 2013, p.8). For our research, we leaned on the objective ending conditions proposed by Nickerson *et al.* (2013): a) a representative sample of objects has been examined, b) no new dimensions or characteristics were added, c) no dimensions or characteristics were split or merged and d) each cell is unique (Nickerson *et al.*, 2013).

Subjective ending conditions, in turn, ensure that a taxonomy is useful. These conditions are met, and the development process is concluded when the taxonomy is deemed concise, robust, comprehensive, extendible, and explanatory (Nickerson *et al.*, 2013). To increase 'robustness' of our subjective ending conditions, we followed Kundisch *et al.'s* (2021) recommendations and evaluated our completed taxonomy. Specifically, we classified NFTs from 25 projects in our taxonomy to ensure that each NFT can be mapped, thus attesting our taxonomy's usefulness (and subjective ending conditions).

| Iteration | Approach | Data | Layers / Dimensions / Characteristics | Status | Ending Conditions |
|---|---|---|---|---|---|
| 1 | C-2-E | Selection of NFT projects identified through a literature review and NFT marketplaces | 4/ 9/ 27 | Initial set of dimensions and characteristics as basis for subsequent iterations | Subjective and objective ending conditions not met: The taxonomy is not explanatory without subject matter knowledge |
| 2 | C-2-E | Exploratory workshop with fellow NFT researchers | 4/ 9/ 29 | Addition of two characteristics within two dimensions | Objective ending conditions not met: Characteristics were added and split, thus requiring a reiteration of the process |
| 3a | E-2-C | NFT expert interviews (academic) | 4/ 13/ 39 | Addition of four dimensions and ten characteristics | Objective ending conditions not met: New dimensions and characteristics were added requiring a reiteration of the process |
| 3b | E-2-C | NFT expert interviews (industry) | 4/ 14/ 42 | Addition of one dimension and three characteristics | All objective and subjective ending conditions are met |

Table I: Taxonomy development process

In terms of design approach, we opted for a mixed approach with three iterations (Table I). We started with a C2E approach that leveraged the existing knowledge on NFTs in the literature (Kundisch *et al.*, 2021; Nickerson *et al.*, 2013; Oberländer *et al.*, 2018). However, as the existing literature on NFTs was still scarce, we also conducted, in a second step, an exploratory workshop with 10 NFT researchers before focusing on an E2C approach in iteration three. In the third iteration, we incorporated the know-how of 11 academic and 15 industry experts (Table II) with whom we conducted 30–60-minute interviews. The experts were located in North America (NA), Europe (EU), Asia (A), and Australia (AUS). The interviews were held in English or German, depending on the language preference of the interviewee. To ensure rigor, we recorded, transcribed, and analyzed the interviews. Where interview partners did not consent to record the sessions, we took extensive notes and transcribed the interview afterwards.

| ID | Position | Relevant Expertise | Location |
|---|---|---|---|
| **Academic Experts** | | | |
| 1 | Senior Researcher | Blockchain, Cryptographic Tokens | EU |
| 2 | Senior Researcher | Blockchain, Cryptographic Tokens | EU |
| 3 | Researcher | Blockchain, NFTs | EU |
| 4 | Researcher | Blockchain, NFTs | EU |
| 5 | Researcher | Blockchain, NFTs | EU |
| 6 | Researcher | Blockchain, NFTs | EU |
| 7 | Researcher | Blockchain, NFT Economics | EU |
| 8 | Researcher | Blockchain, NFTs | EU |
| 9 | Researcher | Blockchain, NFTs | EU |
| 10 | Researcher | Blockchain, NFTs | EU |
| 11 | Researcher | Blockchain, Smart Contracts, NFTs | EU |
| **Industry Experts** | | | |
| 12 | Partner and Member of the Office of the Chief Digital Officer | Digital Assets, NFTs, Blockchain, Metaverse | A |
| 13 | Advisor | Blockchain, NFT Tax & Accounting | EU |
| 14 | Advisor | Blockchain, NFTs | NA |
| 15 | Writer and researcher | Blockchain, NFTs | NA |
| 16 | Attorney | Blockchain, NFTs, Web3 | NA |
| 17 | Artist, Creator and Advisor | NFTs, Media, Design | EU |
| 18 | Project Founder and Technical Product Manager | Blockchain, NFTs | NA |
| 19 | Developer and Technical Cofounder | Blockchain, NFTs, Web3 | EU |
| 20 | Investment Advisor | Blockchain, NFTs, Web3 | EU |
| 21 | Head of Innovation | Blockchain, NFTs, Metaverse | AUS |

| 22 | Deputy Editor in Chief | Blockchain, NFTs | NA |
|----|------------------------|------------------|-----|
| 23 | Senior Project Coordinator for Digital Art Sales | NFTs, Digital Art Sales | NA |
| 24 | Business Developer | Blockchain, NFTs | EU |
| 25 | Founder/ Blockchain Developer | DLT, NFT Licensing | EU |
| 26 | Advisor/ Head of NFT studio | Blockchain, NFTs, Web3 | EU |

Table II: Overview of expert interviews

We selected both the academic and industry experts based on their expertise in the fields of NFTs and DLT more generally. In addition, we required that the interviewed industry experts were involved in NFT projects as a co-founder, investor, or advisor. We used a semi-structured approach to elicit their experience with NFTs (Myers and Newman, 2007). The interviews began with a brief introduction that identified the participating researcher, interviewee, and research project. In the industry expert interviewees, we additionally asked about their involvement in NFT projects, design criteria, and decisions they made during the development of the projects. Based on their answers, we adapted the questions to shift the focus of the interview depending on the interviewees' experience and expertise (Myers and Newman, 2007). In a second step, we then walked the interviewees through the current version of our taxonomy, asking for feedback and concerns.

We conducted the academic and industry expert interviews in batches on a parallel schedule (iterations 3a and 3b in Table II). After conducting batches of 2-4 interviews, we analyzed insights from these interviews and incorporated them into the taxonomy. We continued with these batches until we reached theoretical saturation in the eighth batch, meaning that the interviews in this batch did not generate any new insights and changes. To confirm theoretical saturation, we conducted two more interviews in a ninth batch, which did not yield new results either. In the later batches, we also received clear feedback from our interview partners that our taxonomy is useful, concise, and explanatory.

After the last interview batch, we compiled a broad-ranging set of 25 NFTs from various marketplaces to evaluate our taxonomy. Two co-authors classified the sample independently from one another. This exercise verified the completeness and usefulness of our taxonomy and led us to conclude that it met all our pre-defined ending conditions.

# A Multi-Layer Taxonomy of Non-Fungible Tokens

This chapter presents our multi-layer taxonomy of Non-Fungible Tokens (Table III), which evolved over the three iterations. Our taxonomy includes four layers, 14 dimensions, 42 characteristics, as well as exclusivity markers (ME) to indicate whether the characteristics within a dimension are mutually exclusive (Y) or not (N).

The first layer, 'reference object', describes the object to which the NFT is related. The second layer, 'token properties', describes the technical components of the actual token, whereas 'token distribution' describes how NFTs are issued. Lastly, the 'realizable value' layer sheds light on the value that creators and holders can realize.

## *Reference Object Layer*

The reference object layer is composed of five dimensions, namely "type", "relation", "nature", "adaptability", and "storage".

The **type** describes the underlying reference object of the NFT. The different characteristics of this dimension represent the status quo of the current NFT landscape as identified in our research and confirmed by our expert interviews.

*Artwork* describes a creative item, be it a painting, a video, a song, or any other creative expression commonly referred to as art in its many forms (Davies, 1991; Oxford English Dictionary, 2022b). *Collectibles* represent "items of interest to collectors" (Nadini *et al.*, 2021, p.9). One example for a collectible is an NFT project issued by William Shatner. The actor issued NFT trading cards commemorating 50 moments of his career (shatnercards.io, n.d.), which he intends for his fans to collect.

Further, NFTs can reference *certificates* of various types. Examples include proof-of-attendance or proof-of-authenticity (Zhao and Si, 2021) certificates that provide fraud-proof documentation (Sedlmeir *et al.*, 2022) that a certain task has been accomplished, such as the completion of a training course.

*Permits* such as tickets or passes grant their holder access to certain areas or permission to perform a certain activity. Use cases, which are often referenced, demonstrate the role of NFTs within ticketing systems (Regner *et al.*, 2019) or as ongoing membership passes granting their holders associated benefits (Dwivedi *et al.*, 2022).

NFTs can also represent *accessories*, which are "thing[s that] can be added to something else in order to make it more useful, versatile, or attractive" (Oxford English Dictionary, 2022a). These can range from fashion to gaming items (e.g., wearable items in the Metaverse) (sandbox.game, n.d.-b).

Further, NFTs can represent *real estate* (or any real asset for that matter) in the physical or virtual worlds. For example, virtual land in the latter can be identified by coordinates in the Metaverse (Jeon *et al.*, 2022; Nakavachara and Saengchote, 2022).

Lastly, NFTs can be issued to reference a *domain name*. Domain name NFTs can tie different information, e.g., crypto wallet addresses to a human-readable name (unstoppabledomains.com, n.d.) such as janedoe.crypto.

As certain NFTs represent multiple types at once, we define the dimension as non-exclusive. Additionally, it is important to point out that the list of type characteristics might grow and expand over time, thus making our taxonomy extendible – as required by Nickerson *et al.'s*, (2013) subjective ending conditions. However, all our expert interviews concluded that the characteristics mentioned here are exhaustive (as of writing).

The **relation** domain describes if a reference object is part of a broader context. Reference objects can be *connected* or *unconnected* to other objects of the same or different type. An example for a connected reference object is Bored Ape Yacht Club #8167 (or any other number), as it is part of a collection of 10,000 Bored Ape Yacht Club NFTs. It is therefore *connected*. A stand-alone reference object that is unrelated to any broader context is *unconnected*.

The **nature** distinguishes whether the reference object exists in a *digital* or *physical* form. Digital NFTs include examples such as 'Cryptopunks' (larvalabs.com, n.d.), 'Doodles' (Doodles.app, n.d.) or 'Moonbirds' (moonbirds.xyz, n.d.). They are exclusively confined to the digital realm. In contrast, physical reference objects are those that exist in the 'real world' only. However, to associate an NFT indistinguishably with a physical object, some sort of digital representation or 'intermediate object' is required. Examples include a real-life Rolex watch with an authenticity certificate as an NFT (VIDT Datalink, 2021). Given that a physical reference object remains physical regardless of its digital representation, these characteristics are exclusive.

The **adaptability** dimension differentiates between *static* and *dynamic* reference objects. Static objects cannot be changed, and no attributes can be added or modified after the token has been minted. Dynamic reference objects, in turn, can be modified or changed. As these characteristics are antonyms, we define them as mutually exclusive.

The **storage** dimension distinguishes how NFT reference objects are stored. *On-chain* storage describes that the object is stored on the underlying distributed ledger together with the actual token. *Off-chain* storage refers to a setup where the reference object is stored on a database other than a distributed ledger. Off-chain storage requires the NFT, which remains on-chain, to link to an external storage system for digital or physical objects (Avrilionis and Hardjono, 2022). Both characteristics can also be combined. One could, for example, store a reference object off-chain, generate a hash (a function to generate a unique numerical value off data) of that object, and store that hash on a blockchain. The latter is particularly relevant when it is desired to take physical reference objects into account. Not only would they have to be stored physically, but the storage of their respective digital representation would also need to be weighed following these characteristics. Given the possibility of combining storage options, we classify this dimension as being non-exclusive.

| Layer | Dimension | ME | Characteristic | | | | | | |
|---|---|---|---|---|---|---|---|---|---|
| **Reference Object** | Type | N | Artwork | Collectible | Certificate | Permit | Accessory | Real Estate | Domain Name |
| | Relation | Y | Connected | | | | Unconnected | | |
| | Nature | Y | Digital | | | | Physical | | |
| | Adaptability | Y | Static | | | | Dynamic | | |
| | Storage | N | On-Chain | | | | Off-Chain | | |
| **Token Properties** | Token Standard | Y | Unique Token Standard | | | | Multi-Token Standard | | |
| | Transferability | Y | Restricted | | | | Unrestricted | | |
| | Transparency | Y | Shielded | | | | Unshielded | | |
| | Metadata | Y | Mutable | | | | Immutable | | |
| | Expiration | Y | Defined | | | | Undefined | | |
| **Token Distribution** | Issuance (Minting) | N | Lottery | | Private Sale | | Public Sale | | Airdrop |
| | Schedule | N | One-off | | | Scheduled | | Conditional | |
| **Realizable Value** | Creator | N | Proceeds (Primary Sale) | | Royalties | | Legitimacy | | Loyalty |
| | Holder | N | Proceeds (Resale) | Fees | | Rewards | Utility | Rights | Prestige |

Table III: Multi-layer taxonomy of Non-Fungible Tokens (ME indicates mutual exclusivity of characteristics | Y = mutually exclusive | N = non-exclusive)

## Token Properties Layer

The token properties layer can be broken down into the dimensions 'token standard', 'transferability', 'transparency', 'metadata' and 'expiration'. This layer predominantly describes how the NFT is anchored on a distributed ledger.

The **token standard** can be divided into two characteristics: *unique* or *multi-token standard*. The unique token standard refers – as the name implies – to tokens with only one issue (Ali and Bagui, 2021; Kong and Lin, 2021). Multi-token standards, on the other hand, allow multiple issues of the same token. In addition, some multi-token standards also allow for a combination of fungible and non-fungible tokens (Ali and Bagui, 2021). As NFT creators must choose one standard, this dimension is mutually exclusive.

The **transferability** of NFTs refers to the ability to send a token from one address to another, thus allowing NFTs to be traded and ownership changed. NFTs with *unrestricted* transferability can be moved at will without restriction. *Restricted* NFTs, however, can only be transferred a pre-defined number of times or within a specified timeframe. Some are restricted to zero transfers and remain assigned to a single address (Buterin, 2022; Weyl *et al.*, 2022). Given that these are opposing characteristics, they are mutually exclusive.

The **transparency** dimension describes whether information pertaining to the token is *shielded* or *unshielded*. Unshielded NFTs disclose all information associated with the respective token. Shielded NFTs, however, allow for numerous attributes to be concealed including ownership or transaction history (Babel *et al.*, 2022). While there are different technical options to achieve this, it is essential to note that it is feasible to mint shielded NFTs despite the inherent transparency of DLTs. As a token can only have either of the characteristics, we deem the dimension to be mutually exclusive.

Like the adaptability in the reference object layer, the **metadata** of the token itself can be *mutable* or *immutable*. Immutable metadata cannot be changed after the NFT has been minted: it remains as is. Mutable metadata allows for later changes, an approach which is often used for delayed reveals in NFT projects. Only after the NFT is minted, creators change the metadata to reveal the actual reference object. As this immutability needs to be defined upfront, this dimension remains mutually exclusive.

The dimension **expiration** refers to the technical expiration of an NFT. While it is relatively plausible that a reference object can expire, such as in the case of a certificate or ticket, it does not necessarily affect the associated token. However, tokens can also be *defined* to expire. Like the transparency dimension, there are different technical methods for achieving this (e.g., burn function or attributes 'counting' blocks). Given these binary characteristics, the dimension is mutually exclusive.

## Token Distribution Layer

The token distribution layer describes how the NFT is issued initially. The options are manifold, spanning from lotteries to private and public sales, to airdrop mechanisms. Additionally, there is a timing dimension to the token distribution. It details which schedule the issuance adheres to. This layer focuses on primary issuance only and does not consider secondary distributions, as these are boundless and usually emerge if a token has characteristics of unrestricted transferability.

The **issuance (minting)** dimension describes how NFTs are minted (created). One method is *lotteries* or so-called raffles. NFT lotteries, much like regular lotteries, allow interested users to purchase tickets for a specific NFT. Then, a limited number of tickets are drawn and determined as winners. Once identified, the wallets which hold the winning tickets then qualify for an option to claim the relevant NFT (nftraffles.com, n.d.).

In contrast, a *pre-approved sale* (or an allow-list) refers to a process during which users apply to receive an approval to mint and purchase certain NFTs (binance.com, n.d.) before they become available to the broader public. *Public NFT sales* (public mints), in turn, are open to every user for minting and are typically conducted in a first come, first served manner.

*Airdrops* refer to the subsidized or free distribution of NFTs to users who meet specific, pre-defined criteria (Fröwis and Böhme, 2019), which are set by the creator (Harrigan *et al.*, 2018). Since the distribution of multi-token NFTs can be divided into multiple batches, the issuance dimension is not exclusive.

The **schedule** describes the timetable for distribution. Tokens can be minted in a *one-off* manner or following a pre-defined *schedule*. For instance, in the case of multi-token NFTs, minting can be open for a specific period. Lastly, NFT distribution can be *conditional*, meaning an NFT is only minted once specific criteria are fulfilled. Like the issuance dimension, the schedule dimension is not exclusive.

## Realizable Value Layer

The realizable value layer describes the gain creators and holders of NFTs can realize. These are distinctly different between the two parties and can be of different nature, such as monetary, functional, or social.

NFT **creators** can benefit from the realization of four values. One set of potential value are *proceeds*, which can be generated through *primary sales*. One of the most prominent and lucrative examples is the artwork by artist Beeple, whose NFT "Everydays: The first 5000 days" sold at an auction for USD 69,346,250 (Christie's.com, 2021). Other examples include the sneaker manufacturer Nike, which sold its Metaverse sneakers for over USD 100,000 (Brooks, 2022), or adidas, which sold its "Into the Metaverse" NFTs for USD 22,000,000 (Peters, 2021).

Similarly, creators can also generate *royalties*. Every time an NFT is sold, the creator receives a pre-defined percentage of the resale price – in perpetuity. Additionally, creating and selling NFTs can also lead to an increase in popularity and recognition (artprice.com, 2021), thus, contributing to a creator's *legitimacy* (Vasan *et al.*, 2022).

Lastly, creators stand to increase the interaction with and *loyalty* of their audience or customer base. For example, interviewee 18 pointed out: "Sure, we might generate some revenue with the sale of the NFTs but that is secondary. We are launching our NFT program to deepen customer engagement. As a reward to our loyal customers."

This dimension is not exclusive as multiple values can be realized at once.

NFT **holders** can also realize a variety of values. Following the same line of thinking as articulated previously, holders can generate *proceeds* through the (re)sale of an NFT. A holder can generate *fees* in various forms, for example, by lending an NFT to someone else. An NFT can generate *rewards*, e.g., by staking it and earning so-called staking rewards. Staking describes a process where a holder 'locks' tokens into a platform or on a distributed ledger where it is used for purposes such as trading or consensus finding (a mechanism used to achieve agreement within a DLT network) during the lock-up period. In exchange for providing tokens, the holder receives rewards while maintaining ownership of the token (Bybit.com, 2022).

On a different note, holders can also realize a range of *utility*. An in-game item, for example, can literally be utilized by players of a certain computer game to enable them to perform various tasks and activities. Depending on the type, NFTs can also grant their holders associated *rights*, such as the right to participate in a vote (Ante, 2021). These associated rights, however, can be distinctly different from the rights held by a creator. Creators usually hold the rights associated with the reference object, whereas holders can have separate rights linked to the token. Creators can naturally also be holders and realize values in both dimensions.

Lastly, holding an NFT can also lead to social or cultural *prestige* (Chalmers *et al.*, 2022; Febriandika *et al.*, 2022). As NFTs can be rare and limited editions, owning them indicates exclusivity and cultural prestige (Bateson, n.d.), much like in conventional (art) markets (Kräussl *et al.*, 2016).

As some of these values are only applicable to transferrable tokens and multiple values can be gained simultaneously, this dimension is non-exclusive.

# Evaluation

After completing the development process, we evaluated our NFT taxonomy to demonstrate its usefulness based on a sample of 25 NFTs. Two co-authors conducted the classification independently, concluding that all NFTs could be categorized in our taxonomy. In the following, we demonstrate the evaluation of two selected examples of these 25, representing different reference objects, namely 'real estate', 'artwork', and 'permit'.

## "LAND (192, -156)" NFT

| Layer | Dimension | ME | Characteristic | | | | | | |
|---|---|---|---|---|---|---|---|---|---|
| **Reference Object** | Type | N | Artwork | Collectible | Certificate | Permit | Accessory | Real Estate | Domain Name |
| | Relation | Y | Connected | | | | Unconnected | | |
| | Nature | Y | Digital | | | | Physical | | |
| | Adaptability | Y | Static | | | | Dynamic | | |
| | Storage | N | On-Chain | | | | Off-Chain | | |
| **Token Properties** | Token Standard | Y | Unique Token Standard | | | | Multi-Token Standard | | |
| | Transferability | Y | Restricted | | | | Unrestricted | | |
| | Transparency | Y | Shielded | | | | Unshielded | | |
| | Metadata | Y | Mutable | | | | Immutable | | |
| | Expiration | Y | Defined | | | | Undefined | | |
| **Token Distribution** | Issuance (Minting) | N | Lottery | | Private Sale | | Public Sale | | Airdrop |
| | Schedule | N | One-off | | Scheduled | | Conditional | | |
| **Realizable Value** | Creator | N | Proceeds (Primary Sale) | | Royalties | | Legitimacy | | Loyalty |
| | Holder | N | Proceeds (Resale) | Fees | Rewards | | Utility | Rights | Prestige |

Figure 1: *Classification of "The Sandbox Metaverse LAND (192, -156)" NFT*

'The Sandbox' Metaverse (sandbox.game, n.d.-a) is a Metaverse platform that received wide media attention in 2021 (Marr, 2022). It incorporates various NFT reference objects ranging from accessories to 'real estate'. We will focus on the latter for our first exemplary evaluation (Figure 1).

Overall, the platform offers 166,464 pieces of land, which are identified by their coordinates, thus 'connecting' NFT LAND (192, -156) in a 'digital' map to the broader Sandbox metaverse. Once acquired, attributes (such as the size of the land) cannot be changed, rendering the Sandbox LAND 'static'. The Sandbox uses the InterPlanetary File System (IPFS) to store the reference object 'off-chain' (sandbox.game, n.d.-b). As each plot is clearly identifiable by coordinates, it is issued under a 'unique token standard' making it impossible to issue the same plot of land multiple times.

Much like in the real world, the real estate can be traded on secondary markets and ownership can be traced, thus rendering the token's transferability 'unrestricted' and its transparency 'unshielded'. The token metadata links to a sandbox.game server and is thus 'mutable', while expiration remains 'undefined'.

This particular LAND NFT was issued in an official 'one-off' 'airdrop'. By programming and creating the Metaverse platform, plotting the land and selling it, The Sandbox realized 'proceeds' through a 'primary sale'. Additionally, the platform charges 'royalties' every time the NFT is resold.

As these real estate NFTs can be traded and space is finite, owners can realize 'proceeds' through a 'resale'. Moreover, owners have the 'right' to charge 'fees' for renting out their real estate, e.g., as advertising space. In addition, holders are granted numerous 'rewards', such as airdrops, raffle tickets or exclusive sales. Given that real estate in 'The Sandbox' is partially owned by celebrities (Chirinos, 2022; León, 2022), it puts holders into an exclusive circle of co-owners thus adding to their 'prestige'.

### *"adidas Originals Into the Metaverse (Phase 1)" NFT*

| Layer | Dimension | ME | Characteristic | | | | | | |
|---|---|---|---|---|---|---|---|---|---|
| **Reference Object** | Type | N | Artwork | Collectible | Certificate | Permit | Accessory | Real Estate | Domain Name |
| | Relation | Y | Connected | | | | Unconnected | | |
| | Nature | Y | Digital | | | | Physical | | |
| | Adaptability | Y | Static | | | | Dynamic | | |
| | Storage | N | On-Chain | | | | Off-Chain | | |
| **Token Properties** | Token Standard | Y | Unique Token Standard | | | | Multi-Token Standard | | |
| | Transferability | Y | Restricted | | | | Unrestricted | | |
| | Transparency | Y | Shielded | | | | Unshielded | | |
| | Metadata | Y | Mutable | | | | Immutable | | |
| | Expiration | Y | Defined | | | | Undefined | | |
| **Token Distribution** | Issuance (Minting) | N | Lottery | | Private Sale | | Public Sale | | Airdrop |
| | Schedule | N | One-off | | | Scheduled | | Conditional | |
| **Realizable Value** | Creator | N | Proceeds (Primary Sale) | | Royalties | | Legitimacy | | Loyalty |
| | Holder | N | Proceeds (Resale) | Fees | | Rewards | Utility | Rights | Prestige |

Figure 2: *Classification of the "adidas Originals Into the Metaverse (Phase 1)" NFT*

The apparel and sporting goods manufacturer adidas took a three-phase approach when they entered the NFT space by partnering up with other recognized NFT projects, namely the Bored Ape Yacht Club, PUNKS Comic, and gmoney (adidas.com, n.d.). This collaboration resulted in a visual, artistic representation of the "adidas Originals Into the Metaverse (Phase 1)" NFT (Figure 2) classifying the reference object as 'artwork'. Simultaneously, it is a 'permit' that grants their holders access to exclusive adidas merchandise. The phase 1 NFT is 'connected' to the phase 2 NFT and is thus related to the overall adidas NFT project.

Both objects – artwork and permit – are 'digital' in nature. They are 'static' as no additional attributes (other than the ones initially defined by adidas) can be added to them. A GIF illustrating the NFT is stored 'off-chain' using IPFS. Adidas issued 30,000 phase 1 NFTs in total (adidas.com, n.d.) using a 'multi-token standard'. Transferability is 'unrestricted' as tokens can be exchanged freely.

The creator did not specify any restrictions regarding transparency thus rendering it 'unshielded'. As the token metadata links to an IPFS, it becomes 'immutable'. Phase 1 NFT holders were given the opportunity to burn their token by a certain date and simultaneously mint a phase 2 token, thus 'defining' an expiration for the phase 1 NFT.

Adidas sold 20,000 issues in a 'private sale' for NFT holders of other tokens, such as gmoney tokens, Bored Ape Yacht Club NFTs, etc., thus making the sale 'conditional'. The remaining 10,000 issues were sold in a 'public sale', which concluded by users minting all 30,000 'scheduled' NFTs shortly after the launch. By creating and selling these NFTs, adidas realized 'proceeds' of USD 22,000,000 (Peters, 2021) in 'primary sales'. In addition, adidas is charging 10% in 'royalties' for every future resale. Given this arguably successful launch, adidas received press coverage and 'legitimacy' for being one of the first movers to embrace the Metaverse. In addition, one could argue that the opportunity to participate in the first adidas NFT launch resulted in increased customer 'loyalty'.

Holding this NFT provides 'rights' to claim "physical products designed in collaboration with adidas Originals, Bored Ape Yacht Club, PUNKS Comic and gmoney" (adidas.com, n.d.). All tokens in the project can be traded and thus 'proceeds' from 'resales' can be realized. Given that the number of NFTs is limited, owners stand to gain 'prestige' by holding the tokens (and potentially by owning the exclusive physical products in the real world).

# Discussion

## *Theoretical contribution*

From a theoretical perspective, our multi-layer NFT taxonomy provides an initial, but comprehensive classification of non-fungible tokens. A clear set of layers, dimensions and characteristics makes it possible to understand, distinguish and compare different NFTs. In addition, the taxonomy offers a clear, easy-to-follow nomenclature that future NFT research can build upon. The taxonomy defines what NFTs can represent ('reference object' layer), how they are technically implemented ('token properties' layer), how they are distributed ('token distribution' layer) and which value can be realized by creating or holding them ('realizable value' layer). Via these means, our taxonomy extends current research on decentralized finance and tokenomics, which are still in an exploratory phase. Although prior research classifying cryptographic tokens exists, we found that NFTs are oftentimes mentioned as a mere characteristic. A specific, in-depth understanding of NFTs, however, is missing; a gap that our taxonomy closes. Our taxonomy also takes into account the dynamics of the fast-paced NFT environment and is designed to be extendible, thus allowing new characteristics or dimensions to be added easily.

During our research, we observed that NFTs have implications for different disciplines, including Finance, Marketing, Law, and Information Systems. While NFT research in these areas is growing, we notice that NFTs are often referenced as an umbrella term. Also, NFTs are frequently scoped imprecisely, thus making it difficult to follow which aspects of an NFT are analyzed. We believe our taxonomy can be instrumental for these disciplines in providing a framework that allows for a more precise definition and distinction of NFTs and their characteristics. This will significantly benefit future research in differentiating the aspects under analysis. In addition, our taxonomy can help identify and close gaps in research efforts by raising awareness about dimensions and characteristics that are not obvious at first glance, such as transparency, transferability, metadata, or expiration.

*Practical contribution*

In addition to theoretical contributions, we believe that our taxonomy will also have practical implications for creators, holders, and law makers. The taxonomy can support creators and users to analyze and understand the type of NFT they want to create or acquire. It can help creators to explore and decide which type of NFT to create to fulfill which characteristics. Similarly, holders or buyers of NFTs can use our taxonomy as a guide to analyze which characteristics their desired NFTs hold, and which value they can realize. This allows buyers to make informed decisions before acquisition. Also, as different Metaverses start evolving and spreading throughout the internet, we suppose that the taxonomy can serve as guide to standardize communication and documentation in the form of a concise nomenclature.

We also foresee this taxonomy to find applicability outside of the Metaverse and guide the development of NFT solutions to current, real-world issues. Strengthening property rights (among many other use cases) and making their transfer less bureaucratic and more efficient could potentially be supported by thoughtfully designed NFTs. However, to do so, it will also be necessary to analyze the interplay between rights associated with reference objects and tokens. More generally, given the current discussion around crypto asset regulation, we believe that our taxonomy can inform regulators in their pursuit to regulate NFTs. The EU, for instance, has declared its intention to "prepare a comprehensive assessment [of NFTs]" within 18 months of reaching a provisional agreement in June 2022 of their "markets in crypto-assets (MiCA)" (Council of the European Union, 2022). For this purpose, it might also be useful to analyze and distinguish who issues the NFT. It is probable that regulations might differ from private person to private company to public institution.

# Conclusion

The Metaverse and NFTs are relatively novel concepts that, like many emerging phenomena, are evolving rapidly in unstructured environments. While these circumstances can present significant opportunities, they make it difficult to understand the diversity and potential of NFTs. We close this gap with a multi-layer taxonomy which we developed in a systematic, iterative process following recommendations by Nickerson *et al.* (2013) and Kundisch *et al.* (2021).

As with any research, our work is subject to certain limitations. The NFT industry and NFT research are new, developing fields, which makes it difficult to establish an all-encompassing taxonomy. As NFTs develop, it is probable that our taxonomy might need to be extended. Despite these limitations, our taxonomy is useful for both creators and buyers or holders of NFTs. It helps creators to structure their NFTs to fit their purpose and helps buyers understand what type of NFT they are acquiring.

Moreover, we believe that our taxonomy can support future research in and across different disciplines. Given that NFTs are predominantly tradeable and are partially used as investment vehicles, our taxonomy could be useful in determining strategies for successful NFT investments. It can also help address some of the research questions posed by Dwivedi *et al.* (2022), e.g.: "what are the major drivers for the purchase of NFTs on metaverse[s] by individuals"? Lastly, our taxonomy lays a solid foundation for separating value from hype around NFTs and the Metaverse.